\documentclass[12pt]{article}
\usepackage{epsf,latexsym,amssymb}

\newcommand{\nc}{\newcommand}
\nc{\bea}{\begin{eqnarray}}
\nc{\eea}{\end{eqnarray}}
\nc{\be}{\begin{equation}}
\nc{\ee}{\end{equation}}
\nc{\jo}{J_8}
\nc{\joy}{J_7}
\nc{\imply}{\Rightarrow}
\nc{\alphap}{\alpha^{\prime}}
\nc{\la}{\langle}
\nc{\ra}{\rangle}
\nc{\N}{{\cal N}}
\nc{\WW}{{\cal W}}
\nc{\FF}{{\cal F}}
\nc{\PP}{{\cal P}}
\nc{\M}{{\cal M}_8}
\nc{\Mm}{{\cal M}_3}
\nc{\vp}{\varphi}
\nc{\ox}{\otimes}
\nc{\x}{\times}
\nc{\wt}{\widetilde}
\nc{\Z}{{\bf{Z_2}}}
\nc{\R}{{\cal R}}

\nc{\half}{\frac{1}{2}}
\nc{\tto}{\longrightarrow}
\nc{\dual}{\longleftrightarrow}
\nc{\bc}{\begin{center}}
\nc{\ec}{\end{center}}
\nc{\RR}{{\bf R}}
\nc{\I}{{\cal I}}
\nc{\T}{{\cal T}}
\nc{\ep}{\epsilon}
\nc{\fn}{\footnote}
\nc{\noi}{\noindent}
\nc{\non}{\nonumber}
\nc{\bstar}{\beta^{\ast}}
\nc{\gstar}{\gamma^{\ast}}
\nc{\dstar}{\delta^{\ast}}
\nc{\Tstar}{\Theta^{\ast}}
\nc{\wh}{\widehat}
\nc{\ahat}{\wh{\alpha}}
\nc{\ghat}{\wh{\gamma}}
\nc{\dhat}{\wh{\delta}}
\nc{\That}{\wh{\Theta}}
\nc{\ie}{{\em{i.e. }}}
\nc{\viz}{{\em{viz. }}}
\nc{\ol}{\overline}
\nc{\p}{\partial}
\nc{\w}{\wedge}

\nc{\refb}[1]{(\ref{#1})}

\nc{\sectiono}[1]{\section{#1}\setcounter{equation}{0}}

\begin{document}
\begin{titlepage}
\begin{flushright}
MRI-P-010902\\[2mm]
hep-th/0109076
\end{flushright}

\vskip .6cm

\begin{center}
{\LARGE\bf Type IIA Orientifold Limit of M-Theory\\[3mm] 
on Compact Joyce 8-Manifold of \\[3mm]
$Spin(7)$-Holonomy}
\bigskip 

\vspace*{4.0ex}

{\Large\rm Jaydeep Majumder}\fn{{\rm email} : {\tt 
jaydeep@physics.rutgers.edu,
joydeep@mri.ernet.in}}

\vskip 0.3cm
{\large\it Harish Chandra Research Institute,\\[2mm]
Chhatnag Road, Jhusi,\\[2mm]
Allahabad 211019, INDIA}\fn{Permanent address after October 1, 2001
is {\em New High Energy Theory Center; Rutgers, The State University of
New
Jersey; 126 Frelinghuysen Road; Piscataway, New Jersey
08854-8019}.}

\vskip 2mm
{\rm email} : {\tt jaydeep@physics.rutgers.edu,
joydeep@mri.ernet.in}
\end{center}

\vspace*{5.0ex}

\begin{abstract}

We show that M-theory compactified on a compact Joyce 8-manifold of
$Spin(7)$-holonomy, which yields an effective theory in $D\,=\,3$ with
$\N$ = 1 supersymmetry, admits at some special points in it moduli space a
description in terms of type IIA theory on an orientifold of compact Joyce
7-manifold of $G_2$-holonomy. We find the evidence in favour of this
duality by computing the massless spectra on both M-thory side and type
IIA side. For the latter, we compute the massless spectra by going to the
orbifold limit of the Joyce 7-manifold. 

\end{abstract}

\end{titlepage}

\tableofcontents

\sectiono{Introduction and Summary}\label{s1}

Recently there has been renewed interest in compactification of
string and M-theory on manifolds of exceptional
holonomy\cite{0007213,0011089,0011114,0011130,0101206,0103011,0103115, 
0103167,0103223,
0104105,0104124,0105096,0106026,0106177,0107044,0107177,0108091,0108165,
0108219}. 
These manifolds give rise to models with very low amount of supersymmetry
in spacetime dimensions $D\,\le\,4$.{} Since compactification on these
manifolds preserves very little amount of supersymmetry, they are also
interesting for another reason. Though the string duality conjectures are
now firmly established, there are very few examples where these
conjectures has been tested for $D\,\le\,4$ and $N\,<\,8$ where $N$ is
the total number
of real supercharges present in the theory.  

There are many examples of both compact and non-compact smooth
Riemannian 7-manifolds of $G_2$-holonomy and 8-manifolds of
$Spin(7)$-holonomy. In this paper we shall be mainly concerned
with {\em compact} Riemannian 8-manifold of $Spin(7)$-holonomy\fn{For
{\em non-compact} Riemannian 7-manifolds of $G_2$-holonomy and
8-manifolds of $Spin(7)$-holonomy and string/M-theory dynamics on it, the
reader is referred to the
refs.\cite{0103115,0107177,0108091,bryant-salamon,GPP,0103155,0106034}.}. A
particular orientifold of {\em compact} Riemannian 7-manifold of
$G_2$-holonomy also appears in our work. A large class of {\em compact},
Riemannian 7-manifolds of $G_2$-holonomy and 8-manifolds of
$Spin(7)$-holonomy were constructed by
Joyce\cite{joyce1,joyce2,joyce3}\fn{See also his survey article in
ref.\cite{joyce4} and his book in ref.\cite{joyce5}.}. The construction of
these manifolds is
easier to understand --- Joyce starts with particular 7 or 8 dimensional
compact orbifolds and then the singularities of these orbifolds are blown 
up using
Eguchi-Hanson space\cite{EH} to get a smooth, compact 7- or
8-manifolds of $G_2$- or $Spin(7)$-holonomy respectively\fn{For earlier 
work on string/M-theory on compact Joyce 7- and
8-manifolds, see the
refs.\cite{SV,PT,9603033,9604133,9611036,9812205}. For very recent work
on M-theory on $Spin(7)$-holonomy manifolds, see the ref.\cite{GUKOV}.}.

In this paper we consider M-theory on a particular smooth, compact Joyce
8-manifold of $Spin(7)$-holonomy. We show that if we go to a particular
point in the moduli space of this compactification, the same theory can be
described as an orientifold of type IIA string theory compactified on a
particular Joyce 7-manifold of $G_2$-holonomy. The motivation for this
work came from the work of several authors. In ref.\cite{0103223}, the
authors showed that at a particular loci of moduli space for M-theory
compactified on a compact Joyce 7-manifold of $G_2$-holonomy admits a
description of an orientifold of type IIA string theory compactified on a
Calabi-Yau threefold(its complex dimension is three). In 
ref.\cite{9604133}, the author established a
non-perturbative duality between M-theory compactified on a Joyce
8-manifold of $Spin(7)$-holonomy and heterotic string theory compactified
on a Joyce 7-manifold. We shall try to establish that
heterotic string on a smooth Joyce 7-manifold is dual to an orientifold of
type IIA on the same Joyce 7-manifold, using two methods --- (i) first,
using chain of T- and S-dualities in the orbifold limit and (ii) second,
using the
arguments of fibrewise duality transformation. This is
shown in
subsection \ref{ss5.2}. In ref.\cite{0103115}, the author showed,
using geometrical arguments, M-theory on a manifold of $Spin(7)$-holonomy
can be described locally as a type IIA background where  D6-branes are
wrapped on supersymmetric 4-cycles of a 7-manifold of
$G_2$-holonomy\fn{For an explicit verification of this claim in the 
framework of supergravity, see the ref.\cite{0106055}.}. Although 
he considered non-compact manifolds,
here we shall show explicitly that even for compact cases, D6-branes
appear in type IIA description and wrap a supersymmetric 4-cycle of the
Joyce 7-manifold.

The paper is organised as follows. In section \ref{s2}, we briefly
recapitulate the main properties of compact 
Joyce 8-manifolds of $Spin(7)$-holonomy. In section \ref{s3},
we consider compactification of M-theory on a generic smooth, compact
Joyce 8-manifold of $Spin(7)$-holonomy and write down the massless spectra
obtained in $D\,=\,3$ in terms of Betti numbers of the manifold. In
section \ref{s4}, we choose the simplest of all compact 8-manifolds of
$Spin(7)$-holonomy constructed by Joyce by resolving singularities of a
particular orbifold of 8-torus, $T^8$. Now the Betti numbers of this
smooth Joyce 8-manifold, denoted by $\jo$, can be computed. In section
\ref{s5}, we discuss the type IIA orientifold limit of the M-theory
compactification discussed in section \ref{s4}. We further demonstrate the
connection between heterotic string theory on smooth Joyce 7-manifold
considered by Acharya in ref.\cite{9604133} and the orientifold of type
IIA theory on Joyce 7-manifold we are considering. In section \ref{s6}, we
establish the non-perturbative duality between M-theory on smooth Joyce
8-manifold of $Spin(7)$-holonomy and orientifold of type IIA theory on
Joyce 7-manifold, by computing the massless spectra on type IIA side and
showing its agreement with the computation on M-theory side done in
sections \ref{s3} and \ref{s4}. This computation is actually done by going
over to an equivalent type IIB description of the model and in the
orbifold limit of the smooth Joyce 7-manifold. Next we go back to type IIA
description and show that this theory contains D6-branes wraped on a
supersymmetric 4-cycle inside the compact Joyce 7-manifold. Finally we
show that M-theory on the same compact Joyce 8-manifold of
$Spin(7)$-holonomy admits another orientifold limit of type IIA theory on
Joyce 7-manifold.

\sectiono{Review of Compact $Spin(7)$ Holonomy
Riemannian 8-Manifolds}\label{s2}
%\bigskip

In this section we briefly recall few properties of compact $Spin(7)$
holonomy 8-manifolds. All these materials can be found in the
refs.\cite{bryant-salamon,joyce3,bryant}(See also the chapter 12
of
ref.\cite{salamon} and refs.\cite{GPP,SV,9608116,YO}). We start our
discussion with the concept of $Spin(7)$ holonomy group.
%%%%%%%%%%%%%%%%%%%%%%%%%%%%%%%%%%%%%%%%%%%%%%%%%%%%%%%%%%%%%%%%%%%%%%%%%%%
%\subsection{Preliminaries : $Spin(7)$ Holonomy}
%%%%%%%%%%%%%%%%%%%%%%%%%%%%%%%%%%%%%%%%%%%%%%%%%%%%%%%%%%%%%%%%%%%%%%%%%%%

In Berger's classification\cite{berger} of the possible holonomy groups of
a nonsymmetric, irreducible Riemannian manifold, there are two special
cases, the exceptional holonomy groups $G_2$ in 7 dimensions and $Spin(7)$
in 8 dimensions. We restrict our discussion to $Spin(7)$ holonomy
case. We first discuss the concept of $Spin(7)$-structure on 
a 8-manifold\fn{For more mathematically precise definition, the reader is
referred to the
refs.\cite{bryant-salamon,joyce3,joyce4,bryant,salamon}.}. Let $\M$
be a
compact, Riemannian 8-manifold. Let $\omega$
be a special 4-form on $\M$ which satisfies the following 3 properties :
\begin{enumerate}
\item $\omega$ is {\em closed} as well as {\em co-closed}, \ie
\be\label{e1}
{\rm d}\omega\,=\,0\,= d^{\star}\,\omega
\ee
\item $\omega$ is {\em self-dual} \ie
\be\label{e2}
\omega\,=\,{}^{\ast}\omega
\ee
where ${}^{\ast}$ is the Hodge-star.
\item At each point on $\M$, the {\em stabilizer} of $\omega$ is
isomorphic to the group $Spin(7)\,\subset\,SO(8)$.
\end{enumerate}
Such a 4-form $\omega$ is known as {\em Cayley 4-form}.
Then by an abuse of notation, $\M$ is said to be equipped with a
$Spin(7)$-structure, $\omega$. Since $Spin(7)\,\subset\,SO(8)$, $\omega$
induces a metric $g_{\M}$  and an {\em orientation} on $\M$. Using this 
property of $Spin(7)$ group, $\omega$ can be locally written as
follows\cite{bryant,joyce3}. We
introduce vielbeins\fn{We have suppressed the Lorentz indices of the 
vielbeins for convenience. Thus the index $i$ refers to its tangent space 
indices.} $\{e_i\}$ on a local coordinate chart on $\M$, 
which belongs to the {\em fundamental representation} of $SO(8)$($i$ =
1,$\cdots$,8). In terms of these vielbeins $\omega$ can be written as:
\bea\label{e3}
\omega &=& e_1\w e_2\w e_5\w e_6 \,+\, e_1\w e_2\w e_7\w e_8 \,+\, e_3\w
e_4\w e_5\w e_6 \,+\, e_3\w e_4\w e_7\w e_8 \non\\
&+& e_1\w e_3\w e_5\w e_7 \,-\, e_1\w e_3\w e_6\w e_8 \,-\, e_2\w e_4\w
e_5\w e_7 \,+\, e_2\w e_4\w e_6\w e_8 \non\\
&-& e_1\w e_4\w e_5\w e_8 \,-\, e_1\w e_4\w e_6\w e_7 \,-\, e_2\w e_3\w
e_5\w e_8 \,-\, e_2\w e_3\w e_6\w e_7 \non\\
&+& e_1\w e_2\w e_3\w e_4 \,+\, e_5\w e_6\w e_7\w e_8 
\eea
It is easy to check that this local form of $\omega$ satisfies the
properties given in eqs.\refb{e1} and \refb{e2}. From the definition of
$Spin(7)$ group, it also follows that $\frak{g}\cdot\omega\,=\,\omega$,
$\forall\,\frak{g}\,\in\,Spin(7).$\fn{Note that the volume form of $\M$
\viz
$\omega\w\omega$ gives a natural orientation on $\M$; the metric
on $\M$, $g_{\M}$ is given in terms of vielbeins by
$\displaystyle{\sum_i\,e_i\,\ox\,e_i}$. It can also be proven that $\M$ is
a {\em spin}
manifold\cite{joyce3}.} 

Now we discuss the condition under which $\M$ becomes a $Spin(7)$-holonomy
manifold. Let $\nabla$ be the Levi-Civita connection on $\M$. Then
$\nabla\omega$ is called the {\em torsion} of $\omega$ and $\omega$ is
said to be {\em torsion-free} if 
\be\label{e4}
\nabla\omega\,=\,0
\ee 
It turns out that eq.\refb{e4} is satisfied if and only if eq.\refb{e1}
holds. The necessary and sufficient condition for the holonomy group
$Hol(g_{\M})$ of $g_{\M}$ to be contained in $Spin(7)$ is that $\omega$
should be torsion-free. $Hol(g_{\M})$ may be a proper subgroup of
$Spin(7)$ itself. The necessary and sufficient condition for
$Hol(g_{\M})\,\cong\, Spin(7)$ comes from another topological invariant,
$\wh{A}$-genus, denoted as $\wh{A}(\M)$ of $\M$. It is given by
\be\label{e5}
24\wh{A}(\M)\,=\,-1\,+\,b^1\,-\,b^2\,+\,b^3\,+\,b^4_{+}\,-\,2b^4_{-}
\ee
where $b^k$'s are the Betti numbers of $\M$ \ie
$b^k\,=\,\hbox{dim}\Big(H^k(\M,\RR)\Big)$ and $b^4_{\pm}$ are the
dimensions of the
spaces of self-dual(anti-self-dual) 4-forms in $H^4(\M,\RR)$. It can be
shown that\fn{For a spin manifold like $\M$, $\wh{A}(\M)$ is always an
integer.} $\wh{A}(\M)$ can be either 1, 2, 3 or 4. However\cite{joyce3} 
\be\label{e6}
Hol(g_{\M})\,\cong\,Spin(7)\,, \qquad \hbox{iff} \quad\wh{A}(\M)\,=\,1
\ee
For all other values of $\wh{A}(\M)$, $Hol(g_{\M})$ is a proper subgroup
of $Spin(7).$\fn{The list of all such holonomy groups for all other values
of
$\wh{A}(\M)$ can be found in
ref.\cite{joyce3}.} So if eq.\refb{e5} holds, $\M$ is a compact,
Riemannian $Spin(7)$-holonomy 8-manifold. It also turns out that $\M$ is
{\em simply connected} \ie $b^1(\M)\,=\,0$. Note that, eq.\refb{e6}
further tells that geometrically only 3(not 4) topological numbers of $\M$
are independent.

Thus we learn that for $\M$ to be a $Spin(7)$-holonomy compact 8-manifold,
the Cayley 4-form $\omega$ defined on it should satisfy eq.\refb{e4}; in
other words, it must be {\em covariantly constant}\fn{Compare this
condition with Calabi-Yau $n$-fold. A Calabi-Yau $n$-fold is a Ricci-flat,
compact K\"{a}hler manifold of $SU(n)$-holonomy which is equipped with a
unique,
nonvanishing holomorphic covariantly constant $n$-form(and its
conjugate).}. Now we can use the natural isomorphism between space of
forms and the tensor product of $\Gamma$-matrices of Clifford algebra to
show that the Cayley 4-form on $\M$ is related to a Majorana-Weyl spinor
\fn{In 8 dimensions $\eta$ can be
chosen to be Majorana-Weyl.} $\eta$. In terms of components, we have :
\be\label{e7}
\omega_{\alpha\beta\gamma\delta}\,=\,
\eta^T\,\Gamma_{\alpha\beta\gamma\delta}\,\eta
\ee
As $\omega$ is covariantly constant on $\M$, so also is $\eta$. So we have
a
{\em unique covariantly constant spinor}\fn{Existence of this unique
covariantly constant spinor $\eta$ or a unique zero mode of the Dirac
operator on
$\M$ is actually responsible for the constraint eq.\refb{e5} amongst
the Betti numbers of $\M$ via index theorem\cite{SV,WANG}.} on $\M$
which
will provide upon
compactification with one space-time supersymmetry. Existence of $\zeta$
also automatically shows that $\M$ is {\em Ricci-flat}\cite{salamon}.

In ref.\cite{joyce3}, Joyce constructed compact, smooth,
$Spin(7)$-holonomy Riemannian 8-folds by blowing up the orbifold
$T^8/\Gamma$. Here $T^8$ is the 8-torus which comes with a flat
$Spin(7)$-structure, $\omega_0$ and $\Gamma$ is some finite group of
isometry of $T^8$. Depending upon the nature of the singularities the
resolution differ and sometimes it may not be unique\cite{joyce3}. This
will be discussed in section
\ref{s4}. More details can be found in Joyce's paper\cite{joyce3}. The
particular resolution --- in fact the simplest one --- of $T^8/\Gamma$
which
we shall use in our paper, will be called Joyce 8-manifold and denoted by
$\jo$.

%%%%%%%%%%%%%%%%%%%%%%%%%%%%%%%%%%%%%%%%%%%%%%%%%%%%%%%%%%%%%%%%%%%%%%%%%%
% Let $J_8$ = Spin(7) compact Joyce 8-fold. It's
%nontrivial Betti numbers are\cite{joyce3} :
%\bea
%b^0 &=& b^8 = 1\nonumber\\
%b^2 &=& b^6 \nonumber\\
%b^3 &=& b^5 \nonumber\\
%{\mbox{and}} & & b^{4}_{+},\; b^4_{-}, \;\; {\mbox{where}} \nonumber\\ 
%b^4 &=& b^4_{+} +\; b^4_{-}
%\eea
%where $b^4_{\pm} \to$ dim. of the space of (anti)self-dual 4-forms.
%\noindent{\bf Claim~1}: Dim. of the moduli space of the
%metric preserving the spin(7)-structure on $J_8$ is equal to  $b^4_{-} +
%1$\cite{joyce3}.   
%
%\noindent{\bf Claim ~2}: $J_8$ preserves $1/16$-th. of
%susy. $\exists$ 1 covariantly constant spinor on $\jo$\cite{PT}.
%
%$\imply$ D = 11 SUGRA $\stackrel{on\; \jo}{\tto}$ D = 3, $\N$ = 1 theory.
%%%%%%%%%%%%%%%%%%%%%%%%%%%%%%%%%%%%%%%%%%%%%%%%%%%%%%%%%%%%%%%%%%%%%%%%%%

\medskip
\sectiono{M-Theory on Compact 8-manifold of $Spin(7)$-holonomy and D =
3, $\N$ = 1 Bosonic
Spectra}\label{s3}
%\bigskip

In this section we consider compactification of M-theory on a smooth,
compact Riemannian manifold of $Spin(7)$-holonomy with large volume, so
that we can rely upon supergravity description\fn{Later in section 
\ref{s4}, we compactify on such a  Joyce 8-manifold $J_8$ that the 
topological 
constraints discussed by Sethi, Vafa and Witten\cite{SVW} is avoided. As 
discussed 
in that paper, the tadpole due to the 3-form potential of the 
eleven dimensional supergravity vanishes if the 
Euler number $\chi$ of the 8-manifold is {\em positive} a {\em multiple 
of} 24. In our case 
it turns out that $\chi(J_8) \,=\, 144$. In fact for all the compact 
$Spin(7)$-holonomy 
manifolds constructed by Joyce in ref.\cite{joyce3}, the Euler 
number($\chi$) is 144.}. 
Notice that at low
energies(or long-wavelength) we can use eleven dimensional supergravity
approximation to M-theory. The compactification of eleven dimensional
supergravity theory on smooth, compact 8-manidolds of $Spin(7)$-holonomy
was considered in ref.\cite{PT}. It leads to an effective three
dimensional supergravity theory with $\N$ = 1 supersymmetry. The bosonic
part
of massless spectra of $\N$ = 1, D = 11 supergravity contains graviton,
$G_{MN}$($M$, $N$ = 1,
$\cdots$ ,11) and a 3-form, $A^{(3)}_{MNP}$.\fn{Our notations and
conventions for indices are as follows. We use indices $M$, $N$, $P$,
$\cdots$ for 11-dimensional M-theory. Greek indices like $\mu$, $\nu$
etc. have been used for 10-dimenisonal type IIA or IIB theory. We use the
indices $\bar{\mu}$, $\bar{\nu}$ for efective 3-dimensional theories.}The
non-vanishing
Betti numbers of $Spin(7)$-holonomy Joyce manifolds are $b^2$, $b^3$,
$b^4_{+}$ and $b^4_{-}$. Moreover, the dimension
of the moduli space of metric deformation of a $Spin(7)$-holonomy
Joyce manifold, $\M$ is $(b^4_{-}\,+\,1)$\cite{GPP,joyce3}. Thus the
bosonic part of the massless spectra of the effective theory in $D$ = 
3 with $\N$ = 1 supersymmetry can be determined in the following way: from
the 11-dimensioanl metric, $G_{MN}$  we get a 3-dimensional graviton,
$g_{\mu\nu}$($\mu$, $\nu$ = 1,2,3) and $(b^4_{-} + 1)$
scalars, $\vp_m$, $m$ = 1, $\cdots$ , $(b^4_{-} + 1)$. We also have the
following reduction of $A^{(3)}_{MNP}$ on $\jo$ :
$$
A^{(3)}_{MNP} \sim \sum_{a = 1}^{b^2(\jo)}\,A^a_{\mu}\ox\chi^a_{mn} +
\sum_{i = 1}^{b^3(\jo)}\,S^i \ox \Omega^i_{mnp}\,,
$$
where $\{\chi^a_{mn}\}$ form a basis of $H^2(\jo,\RR)$ and
$\{\Omega^i_{mnp}\}$
form a 
basis of $H^3(\jo,\RR)$. Thus $A^{(3)}_{MNP}$ on $\jo$ gives $b^2$
vectors or $b^2$ scalars\fn{In $D\,=\,3$, a vector is dual to a scalar.},
$\wt{S}^a$; $a = 1$,$\cdots$ ,$b^2$ and
$b^3$ scalars, $S^i$; $i$ = 1, $\cdots$ ,$b^3$. The bosonic part of the
$\N$ =1, D = 3 spectra obtained by compactifying M-theory on a large,
smooth $Spin(7)$-holonomy Joyce 8-fold consists of\cite{PT} {\em one} 
supergravity
multiplet coupled to $(b^2 + b^3 + b^4_{-} + 1)$ scalar multiplets. 

\medskip
\sectiono{A Particular Example of $Spin(7)$ Holonomy Joyce 8-Manifold and
Its Orbifold Limit}\label{s4}
%\bigskip

In ref.\cite{joyce3}, Joyce constructed many $Spin(7)$-holonomy
8-manifolds as blown up orbifolds of 8-torus, $T^8$. Following
Joyce\cite{joyce3}, let us start with $T^8$ and label its coordinates by
$(x^1, \cdots ,x^8)$. Next one has to mod
it out by the
following groups of isometry, $\Gamma$ = $(\Z)^4$ =
$\la\alpha,\beta,\gamma,\delta\ra$ = $\la\alpha, \Theta\ra$\cite{joyce3}
\fn{There exists more general class of $Spin(7)$-holonomy Joyce
8-orbifolds, {\em{viz.}} $T^8/\Z^n$. Here 4 out of $n$ $\Z$ generators
acts non-freely and each reduce the supersymmetry by $\half$. The
remaining $(n \,-\, 4)$ generators act freely and preserve
supersymmetry. Let 
$\N$ = $\#$ susy which remains intact after compactification on $T^8$;
{\em{e.g.}} for M-theory compactified on $T^8$, $\N \,=\, 16$ in D =
3. For
M-theory on $T^8/\Z^4$, we have $\N \,=\, 16/2^4 \,=\, 1$ in D =
3.}. The finite group of isometry so chosen, $\Gamma$ preserves the 
flat $Spin(7)$-structure on $T^8$\cite{joyce3}. Let us now write
down the elements of $\Gamma$ :
\bea\label{Gamma}
\alpha((x^1, \cdots ,x^8)) &=& (-x^1,-x^2, -x^3, -x^4, x^5, x^6, x^7,
x^8)\nonumber\\
\beta((x^1, \cdots ,x^8)) &=& (x^1, x^2,  x^3,  x^4, -x^5, -x^6, -x^7,
-x^8)\nonumber\\
\gamma((x^1, \cdots , x^8)) &=& (c_1 - x^1 , c_2 - x^2, x^3, x^4, c_5 -
x^5, c_6 - x^6, x^7, x^8)\nonumber\\
\delta((x^1, \cdots ,x^8)) &=& (d_1 - x^1, x^2, d_3 - x^3, x^4, d_5 - x^5,
x^6, d_7 - x^7, x^8)\,,
\eea
where $\{c_i\}$ = either 0 or $\frac{1}{2}$; $\{d_j\}$ = either 0 or
$\frac{1}{2}$ $\forall$~ $i$, $j$. It was shown that\cite{joyce3} specific
values of the sets $(c_1,
c_2, c_5, c_6)$ and $(d_1, d_3, d_5, d_7)$ gives rise to 5 types of
singularities. We denote these types of singularities as
{\em{ Type(i)--(v)}}. If one blows up these singularities, one gets a
smooth, compact $Spin(7)$-holonomy Joyce 8-manifold. Each of these
five types of blow ups contribute different numbers of massless scalars
for the M-theory compactification\fn{These massless scalars are
precisely the massless states of various ``twisted sectors'' of
M-thoery compactification on the orbifold $T^8/\Gamma$.}. Out of five 
only {\em three} types of singularities admit unique resolution. Their
contribution to the Betti numbers are as follows\cite{joyce3} :

\noindent{\em{Type(i)}}: Adds 1 to $b^2$, 4 to $b^3$, 3 each to $b^4_{+}$
and $b^4_{-}$.

\noindent{\em{Type(ii)}}: Adds 1 to $b^2$, 3 each to $b^4_{+}$ and
$b^4_{-}$.

\noindent{\em{Type(iii)}}: Adds 1 to $b^4_{+}$.

Each of the other two types of singularities does not admit unique
resolution; each of them gives rise two toplogically distinct resolutions, 
\viz  

\noindent{\em{Type(iv)}} -- Resolution A : Adds 1 each to $b^2$, $b^4_{+}$
and $b^4_{-}$ and 2 to $b^3$.

\noindent{\em{Type(iv)}} -- Resolution B : Adds 2 to $b^3$ and 3 each to
$b^4_{+}$ and $b^4_{-}$.

\noindent{\em{Type(v)}} -- Resolution A : Adds 1 each to $b^2$, $b^4_{+}$
and $b^4_{-}$.

\noindent{\em{Type(v)}} -- Resolution B : Adds 2 each to $b^4_{+}$ and
$b^4_{-}$.

If one choose the sets $(c_1,c_2,c_5,c_6)$ and $(d_1,d_3,d_5,d_7)$
judiciously, one can avoid the singularities of types {\em (iv)} and {\em
(v)} in $T^8/\Gamma$. In this paper we shall choose such values of
these constants that gives rise to unique blow ups for $T^8/\Gamma$. The
simplest choice is\cite{joyce3} :
\be\label{cd}
(c_1, c_2, c_5, c_6) = (\half, \half, \half, \half),\qquad
(d_1, d_3, d_5, d_7) = (0, \half, \half, \half)
\ee
Then the finite group of $\Gamma \cong \langle\,\alpha, \beta, \gamma,
\delta\,\rangle$ is
given by:
\bea\label{Gammafull}
\alpha((x^1, \cdots ,x^8)) &=& (-x^1,-x^2, -x^3, -x^4, x^5, x^6, x^7,  
x^8)\nonumber\\
\beta((x^1, \cdots ,x^8)) &=& (x^1, x^2,  x^3,  x^4, -x^5, -x^6, -x^7,
-x^8)\nonumber\\
\gamma((x^1, \cdots , x^8)) &=& (\half - x^1 , \half - x^2, x^3, x^4,
\half - x^5, \half - x^6, x^7, x^8)\nonumber\\
\delta((x^1, \cdots ,x^8)) &=& ( - x^1, x^2, \half - x^3, x^4,
\half - x^5, x^6, \half - x^7, x^8)
\eea
It has been shown by Joyce that in this case the singular sets of
$T^8/\Gamma$ consists of four singularities of {\em type (i)}, eight
singularities of {\em type (ii)} and sixty four singularities of {\em type
(iii)}. Blowing up all these singularities, we get a smooth, compact
$Spin(7)$-holonomy Joyce 8-manifold $\jo$ with\fn{Note if $(c_1, c_2) \ne
(0,0)$, $(c_5, c_6) \ne
(0,0)$,
$(d_1, d_3) \ne (0,0)$, $(d_5, d_7) \ne (0,0)$ and $(c_1, c_5) \ne (d_1,
d_5)$, then
\bea\label{BettiT8}
b^1(T^8/\Gamma) &=& b^2(T^8/\Gamma) = b^3(T^8/\Gamma) = 0\nonumber\\
b^4_{+}(T^8/\Gamma) &=& b^4_{-}(T^8/\Gamma) = 7\,,
\eea
and $T^8/\Gamma$ is also {\em simply connected}. In fact the only
constant $p$-forms which are invariant under $\Gamma$ are 14 4-forms
appearing in eq.\refb{e3}.} 
\be\label{BettiJ8}
b^2(\jo) = 12,\;\;\; b^3(\jo) = 16,\;\;\; b^4_{-}(\jo) = 43,\;\;\;   
b^4_{+}(\jo) = 107,
\ee
and we shall denote such Joyce 8-manifold as $\jo(12,16,43,107)$. Hence
M-theory on smooth $\jo(12,16,43,107)$ gives an effective $\N = 1$ theory
in $D\,=\,3$ which consists of 1 supergravity multiplet coupled to (12 +
16 + 43 + 1) = 72 scalar multiplets(See also the ref.\cite{9604133}).

\medskip 
\sectiono{IIA Orientifold Limits}\label{s5}

In this section we show that M-theory on $\jo(12,16,43,107)$ has at least
two different type IIA orientifold limits in its moduli space. These
different orientifold limits might be connected to each other by string
dualities.
 
\smallskip 
\subsection{First Orientifold Limit}\label{ss5.1}
%\bigskip

We start our discussion by taking $x^4$ to be the ``M-circle'' or the
compact eleventh dimension and let its radius, $r_4$ be small. As the 
string coupling, $g_{str}$ $\sim$ $r_4^{3/2}$, it implies that we
can use type IIA decription for M-theory compactified on
$\jo(12,16,43,107)$. Thus with $r_4$ small, we can say that M-theory
on $T^8$ is {\em dual to} type IIA theory $T^7$,
where the coordinates on the $T^7$ are $(x^1, x^2, x^3, x^5, x^6, x^7,
x^8)$. Let $\beta^{\ast}$, $\gamma^{\ast}$ and $\delta^{\ast}$ be the
images of $\beta$, $\gamma$ and $\delta$ when the action of the latter are
restricted to $T^7$. They form a group and will be denoted as
$\Tstar\,=\,\la\,\bstar,\gstar,\dstar\,\ra$. Note that none of $\beta$,
$\gamma$
and $\delta$ acts on the M-circle\fn{Note that at this stage, we are not
considering M-theory on $T^8/\Gamma$ but M-theory on
$T^8/\langle\,\beta,
\gamma, \delta\,\rangle$ which preserves $32/2^3$ = 4
supercharges,
giving rise to $\N$ = 2 supersymmetries in D = 3. So type IIA theory on
$T^7/\Tstar$ gives 
$\N$ = 2 supersymmetries in D = 3.} $x^4$. At this point we relabel the
coordinates on $T^7$ as follows :
\bea\label{relabel:comopact}
\hbox{Let}\qquad y^j &=& x^{4 \,+\, j}, \qquad\forall j \,=\, 1, 2, 3, 4
\non\\
y^{4 \,+\, i} &=& x^i, \qquad\forall i \,=\, 1, 2, 3\non\\
{\mbox{and M-circle}}\;\; x^4 &=& y^{10}
\eea
and noncompact coordinates are relabelled as :
\be\label{noncompact}
y^8 \,=\, x^{10},\quad y^9 \,=\, x^9,\quad y^0 \,=\, x^0
\ee
Now we shall specify the
action of the group elements of $\Tstar$ on the coordinates of $T^7$. From
eqs.\refb{Gamma} and \refb{cd}, we find that
\bea\label{Tstar}
\bstar((y^1,\cdots ,y^7)) &=& (-y^1, -y^2, -y^3, -y^4, y^5, y^6, y^7)\non\\
\gstar((y^1,\cdots ,y^7)) &=& (\half - y^1, \half - y^2, y^3, y^4, \half -
y^5, \half - y^6, y^7))\non\\
\dstar((y^1,\cdots ,y^7)) &=& (\half - y^1, y^2, \half - y^3, y^4, -y^5,
y^6, \half - y^7)
\eea
It implies that\fn{This is just an intermediate step.} M-theory on
$T^8/\langle\,\beta, \gamma,
\delta\,\rangle$ with $r_4$ small is {\em dual
to} type IIA theory on \newline $T^7/\langle\,\beta^{\ast}, \gamma^{\ast},
\delta^{\ast}\,\rangle \,\cong\, T^7/\Theta^{\ast}$.
One can now ask the following question: what is the orbifold
$T^7/\Tstar\,$? It is actually the orbifold limit of a
smooth Joyce 7-fold, $J_7(b^2,\, b^3)$ of $G_2$-holonomy\fn{This is a
generalised version of what Joyce discussed in his papers on
$G_2$-holonomy 7-manifolds\cite{joyce1,joyce2}. He took a $\Z^3$ orbifold
of $T^7$,
where the 3 $\Z$ generators  are given as :
\bea
\alpha\Big((x^1, \cdots ,x^7)\Big) &=& (-x^1, -x^2, -x^3, -x^4, x^5, x^6,
x^7)\non\\
\beta\Big((x^, \cdots, x^7)\Big) &=& (b_1 \,-\, x^1, b_2 \,-\, x^2, x^3,
x^4, -x^5, -x^6, x^7)\non\\
\gamma\Big((x^, \cdots, x^7)\Big) &=& (c_1 \,-\, x^1, -x^2, c_3 \,-\, x^3,
x^4, -x^5, x^6, -x^7)\,,
\eea
where the constants, $b_i$ and $c_j$ take values either $0$ or
$\half$. The most general such orbifold can have translations along any of
the inverted directions for each generator. So we need to specify four
constants for each generator and there exists 16 posible
choices for each generator. This leads to huge number of
possibilities. Howeer, string theory on these different backgrounds might
not be totally independent. See ref.\cite{9611036} for further discussion.
%%%%%%%%%%%%%%%%%%%%%%%%%%%%%%%%%%%%%%%%%%%%%%%%%%%%%%%%%%%%%%%%%%%%%%%%%%%
%This gives $16^3$ different orbifolds that we
%can define. I am not sure whether string theory on each these backgrounds
%will be totally independent.
%%%%%%%%%%%%%%%%%%%%%%%%%%%%%%%%%%%%%%%%%%%%%%%%%%%%%%%%%%%%%%%%%%%%%%%%%%%
}\cite{joyce1,joyce2}. To see this, we first have to analyze the
singular sets of fixed points, $S$ of $T^7/\Theta^{\ast}$. It was shown
by Joyce\cite{joyce1,joyce2} that $S$ consists of 12 copies
of $T^3$ and the neighbourhood of each of these singular
submanifolds looks like $T^3\;\ox\;B^4_{\zeta}/\Z$, where $B^4_{\zeta}$ is
a
4-dimensional ball with radius $\zeta$ and $\Z$ acts on it by {\em
inversion} of all of its
four coordinates. This orbifold is blown up by replacing each of
$B^4_{\zeta}/\Z$ by a suitable {\em Eguchi-Hanson} space\cite{EH}. This
operation changes the Betti numbers in the
following way: it adds 1 to $b^2$ and 3 to $b^3$ for each singular
copy\cite{joyce1,joyce2}. Then the orbifold $T^7/\Theta^{\ast}$ is blown
up to a smooth Joyce 7-fold of $G_2$-holonomy, $J_7(b^2,\, b^3)$. Since 
\be\label{betti}
b^1(T^7/\Theta^{\ast}) \;=\;
b^2(T^7/\Theta^{\ast}) \;=\; 0, \qquad
b^3(T^7/\Theta^{\ast}) \;=\; b^4(T^7/\Theta^{\ast})\;=\; 7\,,
\ee
Thus\cite{joyce1,joyce2} 
\be 
b^2(J_7) \;=\; 12, \qquad b^3(J_7) \;=\; 43
\ee
%$\imply$ M-theory on smooth $T^8/\Theta$ with small $x^4$ radius
%$\longleftrightarrow$ IIA on {\em{smooth}} $G_2$-holonomy manifold
%$J_7(12, 43)$, where $\Theta \,\cong\, \langle\,\beta, \gamma,
%\delta\,\rangle$.

We now further mod out $T^8$ by the finite group of isometry $\alpha$ on
M-theory side. This leads to M-theory on $T^8/\la\,\Theta,\alpha\,\ra$ =
$T^8/\Gamma$, where $\Theta \,\cong\, \langle\,\beta, \gamma,
\delta\,\rangle$. Let's consider
the action of $\alpha$ on the coordinates of $T^8$
which 
in turn gives the action of $\alpha$ on the coordinates of $T^7$ and on
the M-circle $y^{10}$ = $x^4$ :
\bea\label{alpha}
\alpha((x^1,\cdots,x^8)) &=& (-x^1, -x^2, -x^3, -x^4, x^5, x^6, x^7,
x^8)\non\\
\alpha((y^1, \cdots ,y^7)) &=& (y^1, y^2, y^3, y^4, -y^5, -y^6, -y^7,
-y^8)\non\\
\alpha((y^{10})) &=& -y^{10}
\eea
It inverts 3 coordinates($x^1$, $x^2$, $x^3$) of $T^8$ and also
the M-circle, $x^4$. In IIA language it corresponds
to\cite{9603113,9707123} 
$\alpha \;\cong\; (-1)^{F_L}\cdot\;\Omega\cdot\R_3$, where $(-1)^{F_L}$
is the image of
the inversion of the M-circle and the factor $\Omega\cdot\,\R_3$ is the
image of
the inversion of the 3 coordinates of $T^8$, in type IIA
theory\fn{Here $\Omega$ is the worldsheet parity transformation and
$\R_3$ denotes inversion of three coordinates of $T^7$.}. If
we blow up the orbifold $T^8/\Gamma$ as done by Joyce\cite{joyce3} and
with the
orbifold group $\Gamma$ is as given in eq.\refb{Gammafull}, we get
M-theory on the smooth $Spin(7)$-holonomy Joyce 8-manifold
$\jo(12,16,43,107)$. According to our ansatz, if we take $r_4$ to be
small, this has a {\em dual} description in terms of type IIA theory on
the orientifold of the {\em smooth} $G_2$-holonomy Joyce 7-fold
$J_7(12,43)$, where the orientifold group is
$\Omega\cdot(-1)^{F_L}\cdot\R_3$. We shall try to verify this claim by
counting and matching the massless spectra on both sides; on M-theory side
we have already computed the massless spectra in section \ref{s4}. In the
next section we
compute the massless spectra on type IIA side on this particular
orientifold background. Before going into this computation, following 
ref.\cite{9611036} we now introduce a new notation
$\I_{lmnp}\,\sigma^l$ where $\I_{lmnp}$
denotes
{\em{inversion}} of the coordinates $(y^l, y^m, y^n, y^p)$ and
$\sigma^l$
denotes half-shifts along the coordiante $y^l$. Thus for example,
\be\label{newnot}
\R_3 \,=\, \I_{567},\quad
\bstar \,=\, \I_{1234},\quad
\gstar \,=\, \I_{1256}\,\sigma^1\,\sigma^2\,\sigma^5\,\sigma^6,\quad
\dstar \,=\, \I_{1357}\,\sigma^1\,\sigma^3\,\sigma^7
\ee
Also $\T_{mnp\cdots}$ will denote T-dualities along $y^m$, $y^n$, $y^p,
\cdots$. Thus according to this new notation, for example, the orbifold
limit of the Joyce 7-manifold $J_7(12,43)$ on which type IIA theory is 
to be compactified, will be denoted as
$T^7/\Big(\I_{1234}, \;\;
\I_{1256}\,\sigma^1\,\sigma^2\,\sigma^5\,\sigma^6, \;\;
\I_{1357}\,\sigma^1\,\sigma^3\,\sigma^7,\;\;  
(-1)^{F_L}\cdot\;\Omega\cdot\;\I_{567}\;\Big)$, which earlier was denoted
as $T^7/\la\,\bstar,\gstar,\dstar\,\ra$.

\subsection{Arguments in Favour of the Proposed Duality}\label{ss5.2}

If the above-mentioned duality conjecture is true \viz for $r_4$
small, M-theory on smooth
$\jo(12, 16, 43, 107)$ is dual to type IIA theory on the orientifold of
$\joy(12,43)$, we
can
verify it at the level of massless spectrum. Since these backgrounds give
rise to very low amount of supersymmetries \viz $\N$ = 1
supersymmetry in D = 3
or, in other words, only two real supercharges, there do not exist any
powerful non-renormalization theorems which otherwise would be able to
protect the result obtained at weak type IIA coupling($g_{\rm
IIA}\,\to\,0$) and extrapolate it to strong coupling($g_{\rm
IIA}\,\to\,\infty$). Nonetheless, we shall give two arguments
in favour of this duality.
%%%%%%%%%%%%%%%%%%%%%%%%%%%%%%%%%%%%%%%%%%%%%%%%%%%%%%%%%%%%%%%%%%%%%%%%%
%using the concept of fibrewise duality
%transformation and a known dual relationship. We know that the 7-fold
%$\joy(12, 43)$ admits associative/coassociative fibrations; \eg it can be
%shown
%that it admits both coassociative $K3$ and $T^4$ fibrations with the same
%3-manifold as the base, say, $\B$. Now we take the known duality \viz
%\be\label{2a-hetdual}
%{\mbox{Type IIA on}}\;\; K3 \;\dual\; {\mbox{Heterotic on}}\;\; T^4
%\ee
%and apply it fibrewise, we find that type IIA on $\joy(12, 43)$ is S-dual
%to
%heterotic on $\joy(12, 43)$. Now we use a known duality relation due to
%Acharya(hep-th/9604133). He showed that heterotic theory on
%smooth $\joy(12, 43)$ is dual to M-theory on $\jo(12, 16, 43, 107)$. So
%finally we can say
%that
%M-theory
%%%%%%%%%%%%%%%%%%%%%%%%%%%%%%%%%%%%%%%%%%%%%%%%%%%%%%%%%%%%%%%%%%%%%%%

\noindent$\bullet\,$ {\em First Argument} :
\medskip

Suppose we consider type IIA
theory on the Joyce 7-manifold $J_7(12,43)$ in its orbifold limit and let
$g_{\rm IIA}\,\to\,\infty$. Now use the following chain of dualities :
%%%%%%%%%%%%%%%%%%%%%%%%%%%%%%%%%%%%%%%%%%%%%%%%%%%%%%%%%%%%%%%%%%%%%%%
%part of which has already
%been
%discussed in eq.\ref{T-dual} below.
%%%%%%%%%%%%%%%%%%%%%%%%%%%%%%%%%%%%%%%%%%%%%%%%%%%%%%%%%%%%%%%%%%%%%%%
%\newpage
\bea\label{dualitychain}
& & {\mbox{ Type IIA on}}\;\; T^7/\Big(\I_{1234}, \;\;
\I_{1256}\,\sigma^1\,\sigma^2\,\sigma^5\,\sigma^6, \;\;
\I_{1357}\,\sigma^1\,\sigma^3\,\sigma^7, \non\\
& & (-1)^{F_L}\cdot\;\Omega\cdot\;\I_{567}\;\Big) \quad
{\mbox{with}}\quad g_{\rm IIA} \,\to\,\infty\nonumber\\ 
&\stackrel{\T_{567}}{\tto}& {\mbox{Type IIB
on}}\;\; \wt{T^7}/\Big(\I_{1234},\;\;
\I_{1256}\,\sigma^1\,\sigma^2\,\sigma^5\,\sigma^6\,((-1)^{F_L})^2,\non\\
& &
\I_{1357}\,\sigma^1\,\sigma^3\,\sigma^7\,((-1)^{F_L})^2,\;\; \Omega\;\Big)
\nonumber\\
&\cong& {\mbox{Type IIB on}}\;\; \wt{T^7}/\Big(\I_{1234},\;\;
\I_{1256}\,\sigma^1\,\sigma^2\,\wh{\sigma^5}\,\wh{\sigma^6},\;\;
\I_{1357}\,\sigma^1\,\sigma^3\,\wh{\sigma^7}, \;\;\Omega\;\Big)
\nonumber\\
&\cong& {\mbox{Type I theory on}}\;\; \wt{T^7}/\Big(\I_{1234},\;\;
\I_{1256}\,\sigma^1\,\sigma^2\,\wh{\sigma^5}\,\wh{\sigma^6},\;\;
\I_{1357}\,\sigma^1\,\sigma^3\,\wh{\sigma^7}\,\Big)\nonumber\\
&\stackrel{{\bf S}}{\tto}& {\mbox{Heterotic
on}}\;\; \wt{T^7}/\Big(\I_{1234},\;\;
\I_{1256}\,\sigma^1\,\sigma^2\,\wh{\sigma^5}\,\wh{\sigma^6},\;\;
\I_{1357}\,\sigma^1\,\sigma^3\,\wh{\sigma^7}\,\Big) \quad
{\mbox{with}} \quad g_{\rm het}\,\to\, 0\nonumber\\
&\stackrel{\T_{567}}{\tto}& {\mbox{Heterotic on
}}\; T^7/\Big(\I_{1234},\;\;
\I_{1256}\,\sigma^1\,\sigma^2\,\sigma^5\,\sigma^6,\;\;
\I_{1357}\,\sigma^1\,\sigma^3\,\sigma^7\,\Big) \quad {\mbox{with}}
\quad g_{\rm het} \,\to\, 0\,,
\eea
where $\wt{T^7}$ denotes $T$-dual of the original 7-torus, $T^7$,
$\wh{\sigma^i}$ denotes {\bf winding shifts} corresponding to the
coordinate $y^i$($i$ = 5, 6, 7), $\T_{567}$ denote T-dualities along
the coordinates $y^5$, $y^6$ and $y^7$ and {\bf S} is the S-duality
map between ten dimensioanl type I theory and heterotic $Spin(32)/\Z$ 
theory. Now we use a known result due to
Acharya\cite{9604133}. He showed that the heterotic string on $\joy(12, 
43)$ is
dual to M-theory on $\jo(12, 16, 43, 107)$. Now look at the last step of
eq.\refb{dualitychain}. The singular 7-manifold on which the heterotic
string is compactified is nothing but the orbifold limit of the smooth
Joyce 7-fold $\joy(12,43)$ of $G_2$-holonomy. Combining his result with
the above chain of
dualities given in eq.\refb{dualitychain}, we get a support for our
conjectured
dual relationship between M-theory on $\jo(12, 16, 43, 107)$ and type IIA
theory on the
particular orientifold of $\joy(12, 43)$. 

\noindent$\bullet\,$ {\em Second Argument} :
\medskip

\begin{figure}[!h]
\begin{center}
\leavevmode
\hbox{%
\epsfxsize=6in
\epsfysize=4.0in
\epsffile{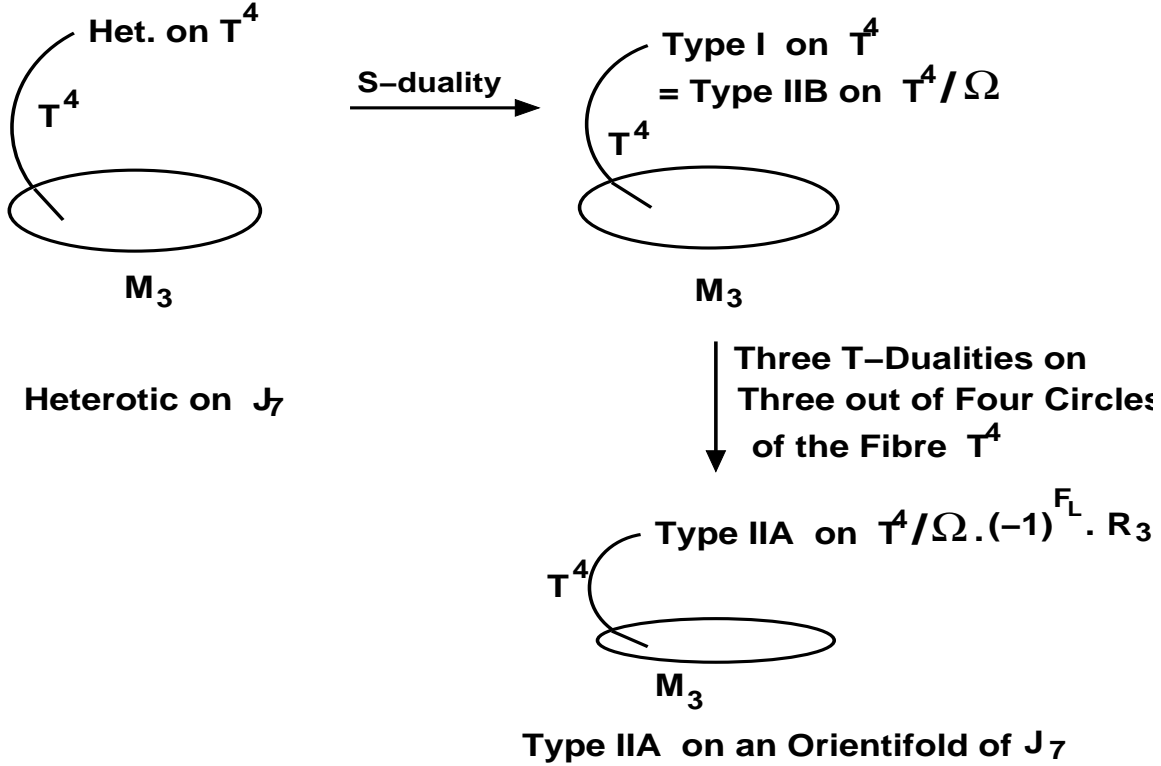}}
%\epsfbox{fig1.eps}
\caption{{\sl Fibrewise Duality Transformation for going over
from Heterotic string theory on smooth Joyce 7-manifold of
$G_2$-holonomy to
type IIA theory on an orientifold of Joyce 7-manifold. We have 
used the fact that the Joyce 7-manifold admits a $T^4$
fibration.}}\label{f1} 
\end{center} 
\end{figure}
We shall give another argument in favour of this conjectured duality using
the concept of fibrewise duality transformation. Joyce 7-manifolds admit
various types of fibrations\cite{9809007}; for example, the manifold under
discussion \viz $J_7(12,43)$ admits a $T^4$ fibration over a particular
base 3-manifold, say, $\Mm$. Let's consider heterotic theory on the fibre
$T^4$. We use the ten dimensional strong-weak duality between heterotic
$Spin(32)/\Z$ and type I $SO(32)$ theories. Using the concept of fibrewise
dualitiy transformation, this leads to type I theory on a manifold $K$
where $K$ admits $T^4$ fibration over a base $\Mm$; see
fig. \ref{f1}. Note that type I theory in ten dimensions can be regarded
as type IIB theory in ten dimensions modded out by $\Omega$. Now we make
$R\,\to\, (1/R)$ duality transformation on three of the four circles of
the fibre $T^4$(See the fig. \ref{f1}). This takes type IIB theory to type
IIA theory. So starting from heterotic string theory on smooth Joyce
7-fold $J_7(12,43)$ which admits $T^4$ fibration and using the concept of
fibrewise duality transformation, we get type IIA theory on the manifold
$\wt{K}$ where $\wt{K}$ is an orientifold of $J_7(12,43)$ and the
orientifold group is $h$ = $\Omega\cdot(-1)^{F_L}\cdot\R_3$.

To verify this duality
relation, in the next section, we compute the massless spectrum of type
IIA theory on this
particular background by going to the orbifold limit of $\joy(12, 43)$.

\medskip 
\sectiono{Calculation of Massless Spectra of Type IIA theory on the
Orientifold of $J_7(12,43)$}\label{s6}

In this section, we compute the spectrum of type IIA theory on the
orientifold of $J_7(12, 43)$ by going to the orbifold limit of
$J_7((12, 43)$. The orbifold of the Joyce 7-manifold $J_7(12,43)$ is given
by $T^7/\Big(\I_{1234}, \;\;
\I_{1256}\,\sigma^1\,\sigma^2\,\sigma^5\,\sigma^6, \;\;
\I_{1357}\,\sigma^1\,\sigma^3\,\sigma^7, 
(-1)^{F_L}\cdot\;\Omega\cdot\;\I_{567}\;\Big)$.
%%%%%%%%%%%%%%%%%%%%%%%%%%%%%%%%%%%%%%%%%%%%%%%%%%%%%%%%%%%%%%%%%%%%%%%%
% where $\Tstar$ $\cong$ $\Z^3$
%is the group of automorphism of $T^7$ and is comprised of 3
%$\Z$ generators, \viz $\bstar$, $\gstar$ and $\dstar$. The action of
%each of these 3 generators on the coordinates of $T^7$ are as follows:
%\bea
%\bstar((y^1,\cdots ,y^7)) &=& (-y^1, -y^2, -y^3, -y^4, y^5, y^6, y^7,
%y^8)\non\\
%\gstar((y^1,\cdots ,y^7)) &=& (\half - y^1, \half - y^2, y^3, y^4, \half
%-
%y^5, \half - y^6, y^7))\non\\
%\dstar((y^1,\cdots ,y^7)) &=& (\half - y^1, y^2, \half - y^3, y^4, -y^5,
%y^6, \half - y^7)
%\eea
%%%%%%%%%%%%%%%%%%%%%%%%%%%%%%%%%%%%%%%%%%%%%%%%%%%%%%%%%%%%%%%%%%%%%%%%
As usual, there will be contribution from the {\em{untwisted}} and
{\em{twisted}} sectors. We are considering type IIA theory on
$T^7/\la\,\Tstar, h\,\ra$ and the massless spectrum in the untwisted
sector is determined by projecting the massless spectrum of 10 
dimensional type IIA theory onto $\Tstar$ invariant subspace of the
full Hilbert space and then projecting these states further onto
$h$-invariant
states. The computation in both untwisted and twisted sectors
are easier if we go over to type IIB description using T-duality\fn{The
spectrum of massless states does not change under T-duality.}. Performing
$T$-duality along $y^1$, we find\fn{Similar $T$-duality was
used in ref.\cite{9611036}.}
\bea\label{T-dual}
& & {\mbox{ Type IIA on}}\;\; T^7/\Big(\I_{1234}, \;\;
\I_{1256}\,\sigma^1\,\sigma^2\,\sigma^5\,\sigma^6, \;\;
\I_{1357}\,\sigma^1\,\sigma^3\,\sigma^7, \non\\
& & (-1)^{F_L}\cdot\;\Omega\cdot\;\I_{567}\;\Big)\nonumber\\
&\stackrel{\T_{1}}{\tto}& {\mbox{Type IIB
on}}\;\; \wh{T^7}/\Big(\I_{1234}\,(-1)^{F_L},\;\;
\I_{1256}\,\wh{\sigma^1}\,\sigma^2\,\sigma^5\,\sigma^6\,(-1)^{F_L}\non\\
& &
\I_{1357}\,\wh{\sigma^1}\,\sigma^3\,\sigma^7\,(-1)^{F_L},\;\; 
\Omega\cdot\I_{1567}\;\Big)
\eea
where $\wh{\sigma^1}$ is the winding shift along $y^1$ as before and
$\wh{T^7}$ denotes that this 7-torus is $T$-dual to the original
one. Henceforth we shall omit the hat on $T^7$.

\noi{\sf Untwisted Sector} 

The spectrum of the untwisted sector of the type IIB theory on the
particular background given in eq.(\ref{T-dual}) can be calculated in two
steps. In the first step we compute the spectrum of type IIB on the
orbifold $J_7$ \ie $T^7/\Big(\I_{1234}\,(-1)^{F_L},\;\;
\I_{1256}\,\wh{\sigma^1}\,\sigma^2\,\sigma^5\,\sigma^6\,(-1)^{F_L},\newline 
\I_{1357}\,\wh{\sigma^1}\,\sigma^3\,\sigma^7\,(-1)^{F_L}\Big)$ $\cong$
$T^7/\wh{\Tstar}$, where $\wh{\Tstar}$ is the image\fn{Similarly the
images of $\bstar$, $\gstar$, $\dstar$ and $h$ under $\T_1$ are
$\wh{\bstar}$, $\wh{\gstar}$, $\wh{\dstar}$ and $\wh{h}$ respectively.} of
$\Tstar$ under
$\T_1$. This can be easily determined if we know the projection rules of
the massless spectrum of 10 dimensional type IIB theory under various 
$\Z$ operators appearing in eq.\refb{T-dual}. The massless bosonic states 
of the type IIB theory in D = 10
consists of a gravtion, $g_{\mu\nu}$,
a 2-form, $B_{\mu\nu}$ and a scalar(dilaton), $\vp$ from the NSNS sector
and a scalar, $a$, another 2-form $B^{\prime}_{\mu\nu}$ and a 4-form,
$C_{\mu\nu\rho\sigma}$(with self-dual 5-form field strength) from the RR
sector($\mu$, $\nu\,=\, 0, 1, \cdots, 9$). If we project these states
onto $\wh{\Tstar}$-invariant subspace, in D = 3 we get 1 vector from the
RR
2-form and 15 scalars altogether --- 8 from the NSNS sector and 7 from 
the RR sector\fn{The 3 dimensional gravtion,
$g_{\bar{\mu}\bar{\nu}}$, antisymmetric 2-forms
$b_{\bar{\mu}\bar{\nu}}$, $b_{\bar{\mu}\bar{\nu}}^{\prime}$ and
their fermionic superpartners obtained in this reduction are all
nondynamical.}. Notice that the $\wh{\Tstar}$ projection preserves
1/8th. of supersymmetries of type IIB theory thus giving rise to $\N$
= 2 supersymmetries in D = 3. Since a {\em vector supermultiplet} of
$\N$ = 2, D = 3 vacuum consists of 1 vector, 1 scalar and
corresponding fermionic superpartners and 
a {\em scalar supermultiplet} of $\N$ = 2, D = 3 vacuum consists of 2
scalars and their fermionic superpartners, then the untwisted sector of
type IIB
on $T^7/\wh{\Tstar}$ gives rise to 1 vector multiplet coupled to 7 scalar
multiplets of $\N$ = 2 supersymmetries in D = 3. In the next step we
project this spectrum
onto $\Omega\cdot\I_{1567}$-invariant sector. This projection further
breaks half of the supersymmetries giving rise to $\N$ = 1 in D = 3, as
expected. It is easy to determine the
$\Omega\cdot\I_{1567}$ projection rules for these states : the vector and
all the 7 scalars from the RR sector do not survive this projection; on
the other hand all the 8 scalars from NSNS sector is even under
$\Omega\cdot\I_{1567}$ and hence survive this projection. So at the end of
the day we find that the untwisted sector of type IIB theory on
$T^7/(\wh{\Tstar},\Omega\cdot\I_{1567})$ consists of 8 scalar multiplets
of $\N$ =1, D = 3 vacuum\fn{Each scalar multiplet of $N$ = 1
supersymmetry in D = 3 consists of a scalar and its fermionic
superpartner. Since a vector in D = 3 is magnetic dual to a scalar, we
need not distinguish between the vector and scalar supermultiplets in D = 
3.}.

\medskip%\newpage
\noi{\sf Twisted Sectors}

The twisted sector states appear at the various fixed points of the
generators of the group $\langle\,\wh{\Tstar},
\,\Omega\cdot\I_{1567}\,\rangle$. So first let us determine the the set of
generators which act non-freely on $T^7$. This set is comprised of
the following 5 elements : 
\begin{enumerate}
\item $\wh{\bstar}$ =
$\I_{1234}\cdot(-1)^{F_L}$,
\item$\wh{\gstar}$ 
= $\I_{1256}\,\wh{\sigma^1}\,\sigma^2\,\sigma^5\,\sigma^6\cdot(-1)^{F_L}$,
\item $\wh{\dstar}$ 
= $\I_{1357}\,\wh{\sigma^1}\,\sigma^3\,\sigma^7\cdot(-1)^{F_L}$,
\item $\wh{h}$ = $\Omega\cdot\I_{1567}$(Image of $h$ under $\T_1$) and
\item$\wh{\bstar}\cdot\wh{h}$
=
$\Omega\cdot(-1)^{F_L}\cdot\I_{234567}$. 
\end{enumerate}
The fixed point manifolds of each of
these elements are actually 3-torus, $T^3$ in $T^7$. Now let us discuss
the twisted
sector states corresponding to each these 5 elements one by one.

If we look at the above list of $\Z$ operators which produce fixed points,
we see that the first three of them have one feature in common --- one of
the factors of these operators is of the form $\I_{pqrs}\cdot(-1)^{F_L}$
for some $p\,\ne\,q\,\ne\,r\,\ne\,s$. The ``twisted'' sector states of
such an operator in type IIB theory consist of
NS5-branes\cite{kutasov}. Let's count the number of fixed points also
: $\wh{\bstar}$ produces 16 fixed points whereas $\wh{\gstar}$ and
$\wh{\dstar}$ produce 8 each\fn{Due to presence of the winding shift 
$\wh{\sigma^1}$, the number of fixed points of each of $\wh{\gstar}$ and
$\wh{\dstar}$ are reduced by half compared to those of
$\wh{\bstar}$.}. We must make sure that we are counting these numbers
on the quotient space not on its any $n$-fold cover($n\,>\,1$). We notice
that the group $\la\wh{\gstar},\wh{\dstar}\ra$ acts freely on the sets of
fixed points of $\wh{\bstar}$, $\wh{h}$ and $\wh{\bstar}\cdot\wh{h}$
whereas the groups $\la\wh{\bstar},\wh{h},\wh{\dstar}\ra$ and
$\la\wh{\bstar},\wh{h},\wh{\gstar}\ra$ act freely on the sets of fixed
points of $\wh{\gstar}$ and $\wh{\dstar}$ respectively. So on the quotient
space $T^7/\wh{\Tstar}$, the set of fixed points of $\wh{\bstar}$
contribute 16/4 = 4 NS5-branes and each set of fixed points of
$\wh{\gstar}$ and $\wh{\dstar}$ contribute only 8/8 = 1 NS5-brane.
Each such NS5-brane wrap different fixed 3-tori and
hence appears to be membrane in 3 dimensions. Henceforth we call such a
membrane as NS-membrane. The massless modes on an NS5-brane of type IIB
theory consists of a vector and
four scalar supermultiplets of D = 6, $\N$ =
(1,1) supersymmetries. Since the vector of this vector supermultiplet
gives rise to 3 scalars upon reduction on $T^3$, thus each NS-membrane
carries 1 vector and 7 scalar supermultiplets of $\N$
= 1 supersymmetries in D = 3 on its worldvoume. But this is not
the end of the story. Notice that
$\wh{h}\,=\,\Omega\cdot\I_{1567}$ fixes the set of fixed points
of both $\wh{\bstar}$ and $\wh{\bstar}\cdot\wh{h}$. So there will be
further appropriate projection of the $\Z$ operator
$\Omega\cdot\I_{1567}$ on the massless modes of the NS5-branes(or
NS-membranes from three dimensional point of view) located at
th fixed points of $\wh{\bstar}$ and on the massless modes of the
D3-branes(or D-membranes from three dimensional point of view) located at
the fixed points of
$\wh{\bstar}\cdot\wh{h}\,=\,\Omega\cdot(-1)^{F_L}\cdot\I_{234567}$. Since
$\wh{h}$ acts freely, as mentioned earlier, on the set of fixed points of
both $\wh{\gstar}$ and $\wh{\dstar}$, there is no further projection on
the massless modes living on the worldvolume of the NS-membranes located
at the fixed points of $\wh{\gstar}$ and $\wh{\dstar}$. The worldvolume
directions of the NS-membranes located at the fixed points of
$\wh{\gstar}$ and $\wh{\dstar}$ are (034789) and
(024689) respectively. There are 2 such NS-membranes and each of them
gives rise to 8 scalar supermultiplets(after dualizing the vector) in 3
dimensions. Thus
NS-membranes located at the fixed points of $\wh{\gstar}$ and
$\wh{\dstar}$ contribute 16 scalar supermultiplets of $\N$ = 1
supersymmetry in D = 3.

\begin{figure}[!h]
\begin{center}
\leavevmode
\hbox{%
\epsfxsize=6.3in
\epsfysize=4.3in
\epsffile{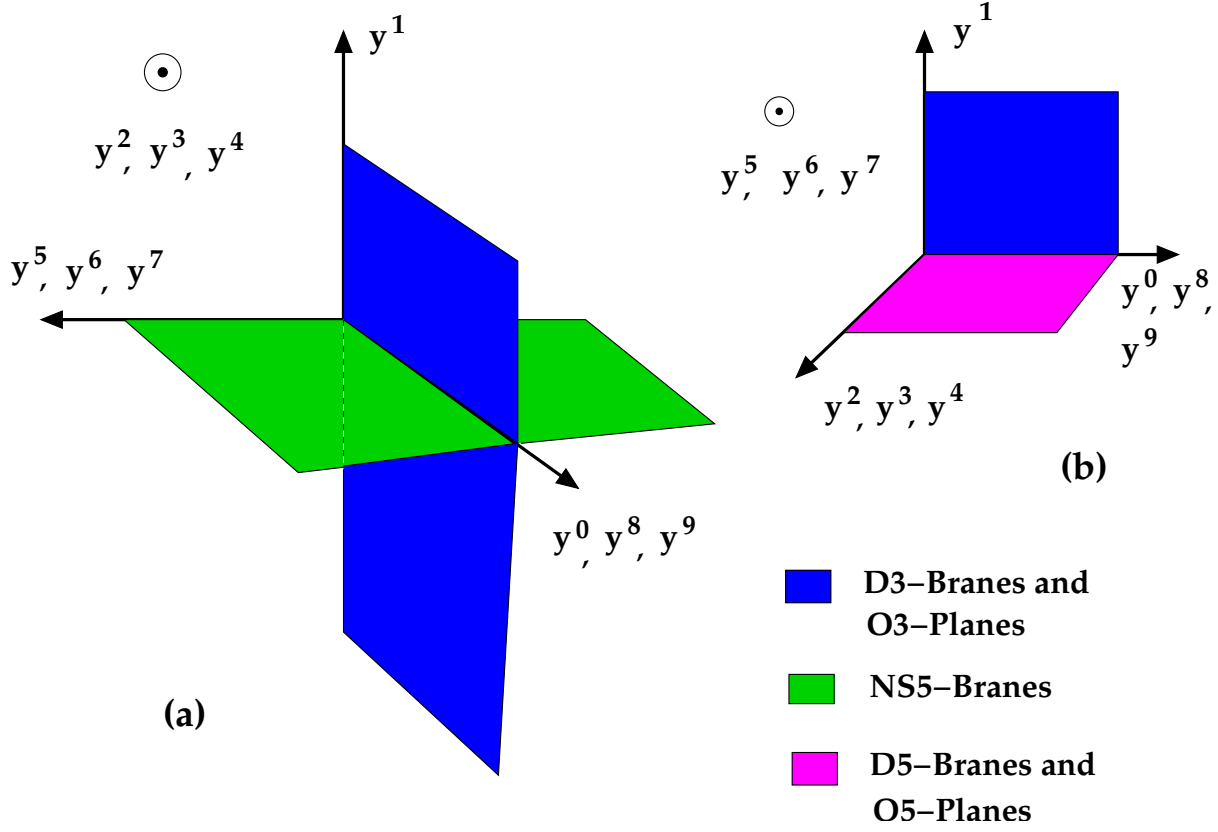}}
%\epsfbox{fig2.eps}
\caption{{\sl Various branes located at different fixed points 
of different ``twisted sectors'' and their intersecting
geometry. In fig. 2(a) we have shown the NS5-branes(green
coloured) located at the fixed points of $\wh{\bstar}$ and
D3-branes plus O3-planes(blue coloured) located at the fixed points
of $\wh{\bstar}\cdot\wh{h}$. Fig. 2(b) shows the intersecting
geometry of these D3-branes plus O3-planes and D5-branes 
plus O5-planes(magenta coloured). The NS5-branes located at
the fixed points of $\wh{\gstar}$ and $\wh{\dstar}$ have not
been shown in the picture.}}\label{f2} 
\end{center} 
\end{figure}
Before counting massless states appearing from twisted sectors of other
fixed points, we look at the geometry of these branes located at these
various fixed
points. These branes intersect amongst themselves and counting of massless
states bocomes easier if we have a picture of this intersecting
geometry. As we already mentioned, we have 4 NS-membranes at the location 
of the fixed points of $\wh{\bstar}$ and 4 D3-branes at the location of
the fixed points of $\wh{\bstar}\cdot\wh{h}$. Notice that we have 16
O3$^{-}$-planes at the fixed points of $\wh{\bstar}\cdot\wh{h}$; the
D3-branes are introduced to cancel the RR charge carried by these
O3$^{-}$-planes\fn{Each O3$^{-}$-plane carries ($-$1/4th.) unit of
D3-brane charge on the quotient space. Henceforth all the charges of 
D-branes or O$p$-planes will be measured on the quotient space.}. The
worldvolume directions of the NS-membranes and D3-branes/O3-planes 
are (056789) and (0189) respectively. Similarly at the location of the
fixed points of
$\wh{h}\,=\,\Omega\cdot\I_{1567}$, we have 4 O5$^{-}$-planes and 4
D5-branes; the D5-branes are introduced to cancel RR charge carried
by the O5$^{-}$-planes\fn{Each O5$^{-}$-plane carries $(-1)$ unit
of D5-brane charge.}. Their common worldvolume directions are
(023489). These
NS-membranes, D3- and D5-branes, O3- and O5-planes intersect as shown in
fig. \ref{f2}.{} The common directions of their worldvolume are the
noncompact directions $y^0$, $y^8$ and $y^9$. The scalars which correspond
to the motion along the transverse coordinate
$y^1$ of the four NS-membranes located at the fixed points of
$\wh{\bstar}$, are {\em odd} under $\wh{h}$-projection. Geometrically it
implies that each of these NS-membranes is stuck to O5-planes but can
move inside this O5-planes. This agrees with the argument given in 
ref.\cite{9703210}. Recall that there exists a vector supermultiplet on
the
worldvolume of an NS5-brane of 10 dimensional type IIB theory. The vector
boson of this multiplet which we denote by $v_{\mu}$($\mu\,=\,0,\cdots,9$) 
gives rise to 3
additional scalars in D = 3. They
are basically the components of this vector boson along the compact
directions $y^5$, $y^6$ and $y^7$ which the NS5-branes of
fig. \ref{f2} wrap. The components of $v_{\mu}$ along noncompact
directions $y^0$, $y^8$ and $y^9$ gives rise to a vector,
$v_{\bar{\mu}}$($\bar{\mu}\,=\,0, 8, 9$) in D = 3. Since
$\wh{h}\,=\,\Omega\cdot\I_{1567}$ fixes the set of fixed points of
$\wh{\bstar}$, we need to determine the projection rules of
$\Omega\cdot\I_{1567}$ on various components of $v_{\mu}$. The low-energy
effective action of the massless modes of type IIB
NS5-brane includes a term\cite{EYRAS-JANSSEN-LOZANO}
\be\label{NS5}
\int_{\WW_6}\,\PP\big[C^{(4)}\big]\wedge\FF\,,
\ee
where
\be\label{FF}
\FF\,=\,{\rm d}v \,+\, \frac{1}{2\pi\alphap}C^{(2)}
\ee
Here $C^{(2)}$ and $C^{(4)}$ are RR 2-form and 4-form fields of type IIB
theory respectively, $\PP\big[C^{(4)}\big]$ denotes the pullback of the
4-form potential $C^{(4)}$ on the worldvolume of NS5-brane and $\WW_6$
denotes the worldvolume
of an NS5-brane. From eq.\refb{NS5}, we can easily find that $v_5$,
$v_6$, $v_7$ are {\em odd} whereas the components $v_0$,
$v_8$ and $v_9$ are {\em even} under $\Omega\cdot\I_{1567}$.
Thus all of the four NS-membranes at the fixed points of $\wh{\bstar}$
contributes 4 vector supermultiplets or after dualizing the vector, 4
scalar supermultiplets in D = 3. 

We now turn to the ``twisted sector'' states located at the fixed points 
of
$\wh{\bstar}\cdot\wh{h}$. As already mentioned, such fixed points on
the quotient space give
rise to 16 O3-planes and 4 D3-branes wrapped along $y^1$. So these
D3-branes appear to be D-membranes in 3 dimensions. These D3-branes cut
O5-planes
stretched along (023489) directions, orthogonally. This implies that
there will be further projection on the massless fields living on
D3-branes' worldvolume. In fact this can be understood as follows: since
$\wh{h}$ fixes these set of fixed points, we need to project the massless
states living on the worldvolume of D3-branes onto
$\Omega\cdot\I_{1567}$-invariant sector. The bosonic contents of massless
fields living on
such
a D3-brane from 3 dimensional point of view are 6 transverse scalars,
$y^2$, $y^3$, $y^4$, $y^5$, $y^6$ and
$y^7$, the component of the vector along compact worldvolume direction
$y^1$ which is counted as a scalar in D = 3 and denoted as $\wt{A_1}$ and
a
vector, $\wt{A_{\bar{\mu}}}$; $\bar{\mu}\,=\,0, 8, 9$. The 3
dimensional vector $\wt{A_{\bar{\mu}}}$ and 3 transverse scalars
corresponding to the movement
along $y^5$, $y^6$ and $y^7$ can be shown to be {\em
odd} under $\Omega\cdot\I_{1567}$. Geometrically it implies that the
D3-branes are stuck to the O5-planes. The scalars corresponding to the
movement along $y^2$, $y^3$ and $y^4$ and $\wt{A_1}$ are {\em even} under
$\Omega\cdot\I_{1567}$ projection. So finally the D3-branes located at the
fixed points of $\wh{\bstar}\cdot\wh{h}$ contribute $4\x 4$ = 16 scalar
supermultiplets of $\N$ = 1 supersymmetry in D = 3.
 
The ``twisted sector'' states corresponding to the element
$\wh{h}$ = $\Omega\cdot\I_{1567}$ consists of 4 O5$^{-}$-planes and 4
D5-branes on the quotient space, wrapping the directions $y^2$, $y^3$ and
$y^4$ 
of the orbifold. The common worldvolume directions of O5-planes and
D5-branes are (023489) as shown in fig. \ref{f2}. From the 3-dimensional
point of view, these D5-branes are D-membranes. The bosonic contents of
the massless fields living on such a D5-brane or D-membrane from three
dimensional point of view, consists of 4 transverse scalars : $y^1$,
$y^5$, $y^6$ and $y^7$, three massless scalars $A_2$, $A_3$ and $A_4$
corresponding to the components of the vector on a D5-brane along the
compact coordinates $y^2$, $y^3$ and $y^4$ respectively and a three
dimensional vector, $A_{\bar{\mu}}$; $\bar{\mu}\,=\,0, 8, 9$. Now notice
that $\wh{\bstar}\,=\,\I_{1234}\cdot(-1)^{F_L}$ fixes the set of fixed
points of $\wh{h}$ = $\Omega\cdot\I_{1567}$. Hence there will be further
projection due to $\wh{\bstar}$ on the massless modes of D5-branes located
at the fixed points of $\wh{h}$. The transverse scalars $y^5$, $y^6$ and
$y^7$ and the three dimensional vector $A_{\bar{\mu}}$ turn out to be
{\em even} under $\I_{1234}\cdot(-1)^{F_L}$. The massless scalars $A_2$,
$A_3$ and $A_4$ and the transverse scalar $y^1$ are
$\I_{1234}\cdot(-1)^{F_L}$ {\em odd}. Thus after dualizing the vector to a
scalar, the D5-branes located at the fixed points of $\wh{h}$ contribute
$4\,\times\,4$ = 16 scalar supermultiplets in $D\,=\,3$ of $\N$ = 1
vacuum.

Summing up the contribution of all five ``twisted sectors'', we find
that they contribute a total of $4\,\times\,16$ = 64 scalar
supermultiplets of $\N$ = 1 supersymmetry in $D\,=\,3$. If
we add up the contribution of the untwisted and twisted sectors, we see
that massless spectra of type IIB on the particular orbifold
under consideration consists of one supergravity multiplet
coupled to ($8\,+\,64$) = 72 scalar multiplets of $\N$ = 1
supersymmetry in D = 3. As $T$-duality does not change the
massless spectra, we make an inverse
$T$-duality($\T_1^{-1}$) along $y^1$ to go back to the type
IIA background $T^7/(\Tstar, h)$ we started with. So the massless
spectra of type IIA theory on $T^7/(\Tstar, h)$ consists of one
supergravity
multiplet coupled to 72 scalar multiplets of $\N$ = 1
supersymmetry
in D = 3. Since $T^7/(\Tstar, h)$ is actually an orientifold of smooth 
Joyce 7-manifold $J_7(12,43)$ in the orbifold limit, this result agrees
with the
result of compactification of M-theory on the smooth Joyce 8-fold, $\jo$
derived in section \ref{s4}. Agreement of the massless spectra gives a
support in favour of our strong-weak duality conjecture. 

\medskip
\subsection{Appearance of D6-branes in IIA Description}\label{ss6.1}

According to ref.\cite{0103115}, if D6-branes of type IIA theory wraps a
supersymmetric 4-cycle of a $G_2$-holonomy 7-manifold, its lift to
M-theory is locally described by M-theory on a smooth 8-manifold of
$Spin(7)$-holonomy. In this paper we are considering M-theory on a smooth
$Spin(7)$-holonomy Joyce 8-manifold and find that at some particular
point in its moduli space, it can be described as an orientifold of type
IIA on smooth Joyce 7-manifold of $G_2$-holonomy. Naturally one might ask
the following question: where are the D6-branes in type IIA
description? What kind of 4-cycles do they wrap? The answer to the first
question is: yes, the D6-branes are indeed present in type IIA
description. Note that in the equivalent type IIB description as
discussed in the last section, we have
found in one of the ``twisted sectors'', D5-branes wrapping certain three
cycles of the orbifold. As we perform an inverse $T$-duality along $y^1$
to get back to the type IIA description, they become D6-branes wrapped on
certain 4-cycles which in the orbifold limit are represented as
$T^4/\Z$. Once we blow up the orbifold $T^7/\Tstar$ to get the smooth
Joyce 7-manifold $J_7(12,43)$, these D6-branes actually
wrap a
supersymmetric coassociative
4-cycle\footnote{Here $\frak{N}$  is a K3 surface.}, $\frak{N}$ $\in\,
J_7(12,43)$. Notice that these coassociative 4-cycles are not {\em rigid},
as the D6-branes wrapped on it have 3 moduli corresponding to their
movement along the compact directions $y^5$, $y^6$ and $y^7$.\fn{This
agrees
with Mclean's Theorem\cite{mclean} which states that moduli space of 
coassociative submanifold, $\frak{N}$ is locally a smooth manifold and its
dimension is equal to $b^2_{+}(\frak{N})$ = \# Self-dual 2-forms on
$\frak{N}$. In
this case $\frak{N}$ $\cong$ K3 and thus $b^2_{+}(\frak{N})$ =
$b^2_{+}$(K3) = 3.}

\medskip
\subsection{Second Orientifold Limit}\label{ss6.3}

In this section, we briefly mention that M-theory on the particular
$Spin(7)$-holonomy Joyce 8-manifold $\jo(12,16,43,107)$ at some particular
point in its moduli space admits another type IIA description on some
orientifold of smooth Joyce 7-fold of $G_2$-holonomy. This time we let
$x^8$ be the ``M-circle'' and $r_8$ be small; in other words, 
$g_{str}$
$\sim$ $r_8^{3/2}$ is also small.. Note that none of $\alpha$, $\gamma$
and $\delta$ of
eq.\refb{Gammafull} acts on
$x^8$. So if we let $\ahat$, $\ghat$ and $\dhat$ be the
action of $\alpha$, $\gamma$ and $\delta$ restricted to $T^7$, then we can
show that $T^7$/($\ahat$, $\ghat$, $\dhat$) $\cong$ $T^7$/$\That$ is a
Joyce 7-fold orbifold and
it has a unique resolution to a smooth Joyce 7-fold $J_7$ with $b^2(J_7)$
= 12, $b^3(J_7)$ = 43. One can show that in a similar fashion that the
non-identity elements of $\That$ which act non-freely on $T^7$ are
$\ahat$, $\ghat$ and $\dhat$. The singular set $\wh{S}$ in $T^7$/$\That$
which is
a union of all the singular sets of the operators $\ahat$, $\ghat$ and
$\dhat$ is given by 12 copies of 3-tori, $T^3$. Repeating our arguments of
section \ref{s5}, we can
show that $\beta$ of eq.\refb{Gammafull} in the type IIA picture can be
identified with $(-1)^{F_L}\cdot\Omega\cdot\I_{567}$. Thus
M-theory on the same smooth Joyce 8-fold $\jo(12,16,43,107)$ in the limit
of {\em{small}} $r_8$, is dual to type IIA
on $J_7(12, 43)/(-1)^{F_L}\cdot\Omega\cdot\R_3$.

Let's call this second orientifold background of type IIA theory be
``orientifold II'' and that of subsection \ref{ss5.1} be ``orientifold
I''. Notice that the role of the dilaton in orientifold I is played by a
geometrical modulus in orientifold II and vice versa, though they are
actually compactifications on the same target space. This strongly
indicates that these two orientifold pictures \viz orientifold I and
orientifold II of subsections \ref{ss5.1} and \ref{ss6.3} are related by
some kind of duality relations, although its precise nature is not known
to the author.

\bigskip
\begin{center}
{\large\bf Acknowledgement}
\end{center}
\medskip

I would like to thank Ashoke Sen who read a preliminary draft of this
paper carefully and discussed many points of the paper at various stages
of this work. I would also like to thank Debashis Ghoshal, Dileep Jatkar,
Satchi
Naik and Sudhakar Panda for illuminating discussion. An email
correspondence by Bobby Acharya is gratefully acknowledged. 

% Bibliography for my paper

\end{document}